\documentclass[10pt,twocolumn,conf]{IEEEtran}
\usepackage{times}
\usepackage{amsbsy}
\usepackage{latexsym}
\usepackage{amsmath}
\usepackage[ansinew]{inputenc}
\usepackage[dvips ]{graphicx}
\input epsf
\usepackage{epic,eepic}
\usepackage{url}

\IEEEoverridecommandlockouts

\title{\Huge Interference Suppression and Group-Based Power Adjustment via Alternating Optimization
for DS-CDMA Networks with Multihop Relaying \vspace{0.05em}}
\author{Rodrigo C. de Lamare  \\ Communications Research Group \\ Department of Electronics,
    University of York, York Y010 5DD, United Kingdom \\
    Emails: \protect\url{rcdl500@ohm.york.ac.uk}
\thanks{\footnotesize The work of the author was
supported by the University of York, York Y010 5DD, United
Kingdom.   }}

\begin{document}
\maketitle

\begin{abstract}

This work presents joint interference suppression and power
allocation algorithms for DS-CDMA networks with multiple hops and
decode-and-forward (DF) protocols. A scheme for joint allocation
of power levels across the relays subject to group-based power
constraints and the design of linear receivers for interference
suppression is proposed. A constrained minimum mean-squared error
(MMSE) design for the receive filters and the power allocation
vectors is devised along with an MMSE channel estimator. In order
to solve the proposed optimization efficiently, a method to form
an effective group of users and an alternating optimization
strategy are devised with recursive alternating least squares
(RALS) algorithms for estimating the parameters of the receiver,
the power allocation and the channels. Simulations show that the
proposed algorithms obtain significant gains in capacity and
performance over existing schemes.

\end{abstract}

\section{Introduction}

Multiple-antenna wireless communication systems can exploit the
spatial diversity in wireless channels, mitigating the effects of
fading and enhancing their performance and capacity. Due to the
size and cost of mobile terminals, it is considered impractical to
equip them with multiple antennas. However, spatial diversity
gains can be obtained when single-antenna terminals establish a
distributed antenna array via cooperation
\cite{sendonaris}-\cite{laneman04}. This allows a significant
reduction on the transmitted power for an equivalent performance.
In a cooperative system, terminals or users relay signals to each
other in order to propagate redundant copies of the same signals
to the destination user or terminal. To this end, the designer
must use a cooperation protocol such as amplify-and-forward (AF)
\cite{laneman04}, decode-and-forward (DF) \cite{laneman04,huang}
and compress-and-forward (CF) \cite{kramer}.

Recent contributions in the area of cooperative and multihop
communications have considered the problem of resource allocation
\cite{luo,long}. Prior work on cooperative multiuser DS-CDMA
networks has focused on the assessment of the impact of multiple
access interference (MAI) and intersymbol interference (ISI), the
problem of partner selection \cite{huang,venturino}, the bit error
ratio (BER) and outage performance analysis \cite{vardhe}, and
training-based joint power allocation and interference mitigation
strategies \cite{delamare_jpais,joung}. However, these strategies
require a higher computational cost to implement the power
allocation and a significant amount of signalling, decreasing the
spectral efficiency of cooperative networks. This problem is
central to ad-hoc and sensor networks \cite{souryal} that utilize
spread spectrum systems and require multiple hops to communicate
with nodes that are far from the source node.

In this work, joint interference suppression and power allocation
algorithms for DS-CDMA networks with multiple hops and DF
protocols are proposed. A scheme that jointly considers the power
allocation across the relays subject to group-based power
constraints and the design of linear receivers for interference
suppression is proposed.  The idea of a group-based power
allocation constraint is shown to yield close to optimal
performance, while keeping the signalling and complexity
requirements low. A constrained minimum mean-squared error (MMSE)
design for the receive filters and the power allocation vectors is
developed along with an MMSE channel estimator for the cooperative
system under consideration. The linear MMSE receiver design is
adopted due to its mathematical tractability and good performance.
However, the incorporation of more sophisticated detection
strategies including interference cancellation with iterative
decoding \cite{delamare_mber}-\cite{delamaretc} and advanced parameter estimation
methods \cite{delamaresp}-\cite{jidf} are also possible. In order to solve the
proposed optimization problem efficiently, a method to form an
effective group of users and an alternating optimization strategy
are presented with recursive alternating least squares (RALS)
algorithms for estimating the parameters of the receiver, the
power allocation and the channels.

The paper is organized as follows. Section II describes a
cooperative DS-CDMA system model with multiple hops. Section III
formulates the problem, details the constrained MMSE design of the
receive filters and the power allocation vectors subject to a
group-based power allocation constraint, and describes an MMSE
channel estimator. Section IV presents an algorithm to form the
group and the alternating optimization strategy along with
RLS-type algorithms for estimating the parameters of the receiver,
the power allocation and the channels. Section V presents and
discusses the simulation results and Section VI draws the
conclusions of this work.

\section{Cooperative DS-CDMA Network Model}

Consider a synchronous DS-CDMA network with multipath channels,
QPSK modulation, $K$ users, $N$ chips per symbol and $L$ as the
maximum number of propagation paths for each link. The network is
equipped with a DF protocol that allows communication in multiple
hops using $n_r$ fixed relays in a repetitive fashion. We assume
that the source node or terminal transmits data organized in
packets with $P$ symbols, there is enough control data to
coordinate transmissions and cooperation, and the linear receivers
at the relay and destination terminals are synchronized with their
desired signals. The received signals are filtered by a matched
filter, sampled at chip rate and organized into $M \times 1$
vectors ${\boldsymbol r}_{sd}[m_j]$, ${\boldsymbol r}_{sr_i}[m_j]$
and ${\boldsymbol r}_{r_id}[m_j]$, which describe the signal
received from the source to the destination, the source to the
relays, and the relays to the destination, respectively,
\begin{equation}
\begin{split}
{\boldsymbol r}_{sd}[m_1] & = \sum_{k=1}^K  a_{sd}^k[m_1]
{\boldsymbol C}_k {\boldsymbol h}_{sd,k}[m_1]b_k[m_1]   +
{\boldsymbol \eta}_{sd}[m_1] \\ & \quad  + {\boldsymbol
n}_{sd}[m_1],
\\
{\boldsymbol r}_{sr_j}[m_1] & = \sum_{k=1}^K a_{sr_j}^k[m_1]
{\boldsymbol C}_k {\boldsymbol h}_{sr_j,k}[m_1] {b}_k[m_1]  +
{\boldsymbol \eta}_{sr_j}[m_1] \\ & \quad + {\boldsymbol
n}_{sr_j}[m_1],
\\
{\boldsymbol r}_{r_jd}[m_j] & = \sum_{k=1}^K a_{r_jd}^k[m_j]
{\boldsymbol C}_k {\boldsymbol h}_{r_jd,k}[m_j] \tilde{b}_k[m_j]  +
{\boldsymbol \eta}_{r_jd}[m_j] \\ &  \quad+ {\boldsymbol
n}_{r_jd}[m_j],
\\ j & = 1, \ldots,n_p,~  m_j  = (j-1)P+1, \ldots, jP,\\
 i& =1, \ldots, P \label{rvec}
\end{split}
\end{equation}
where $M=N+L-1$, $P$ is the number of packet symbols, $n_p=n_r+1$ is
the number of transmission phases or hops, $n_r$ is the number of
relays, and $m_j$ is the index of original and relayed signals. The
vectors ${\boldsymbol n}_{sd}[m_1]$, ${\boldsymbol n}_{sr_j}[m_1]$
and ${\boldsymbol n}_{r_jd}[m_j]$ are zero mean complex Gaussian
vectors with variance $\sigma^2$ generated at the receivers of the
destination and the relays from different links, and the vectors
${\boldsymbol \eta}_{sd}[m_1]$, ${\boldsymbol \eta}_{sr_j}[m_1]$ and
${\boldsymbol \eta}_{r_jd}[m_j]$ represent the intersymbol
interference (ISI). The amplitudes of the source to destination,
source to relay and relay to destination  links for user $k$ are
denoted by $a_{sd}^k[m_1]$, $a_{sr_j}^k[m_1]$ and $a_{r_jd}^k[m_j]$,
respectively. The quantities ${b}_k[m_1]$ and $\tilde{b}_k[m_j]$
represent the original and reconstructed symbols by the DF protocol
at the relays, respectively. The $M \times L$ matrix ${\boldsymbol
C}_k$ contains versions of the signature sequences of each user
shifted down by one position at each column as described by
\begin{equation}
{\boldsymbol C}_k = \left[\begin{array}{c c c }
c_{k}(1) &  & {\bf 0} \\
\vdots & \ddots & c_{k}(1)  \\
c_{k}(N) &  & \vdots \\
{\bf 0} & \ddots & c_{k}(N)  \\
 \end{array}\right],
\end{equation}
where ${\boldsymbol c}_k = \big[c_{k}(1), ~c_{k}(2),~ \ldots,~
c_{k}(N) \big]$ stands for the signature sequence of user $k$, the
$L \times 1$ channel vectors  from source to destination, source to
relay, and relay to destination are ${\boldsymbol h}_{sd,k}[m_1]$,
${\boldsymbol h}_{sr_j,k}[m_1]$, ${\boldsymbol h}_{r_jd,k}[m_j]$,
respectively. By collecting the data vectors in (\ref{rvec})
(including the links from relays to the destination) into a
$(n_r+1)M \times 1$ received vector at the destination we obtain {
\begin{equation}
\begin{split}
 \hspace{-0.5em}\left[\begin{array}{c} \hspace{-0.5em}
  {\boldsymbol r}_{sd}[m_1] \\
 \hspace{-0.5em} {\boldsymbol r}_{r_{1}d}[m_2] \\
\hspace{-0.5em}  \vdots \\
 \hspace{-0.5em} {\boldsymbol r}_{r_{n_r}d}[m_{n_p}]
\end{array}\right] & = \left[\begin{array}{c}
  \sum_{k=1}^K  a_{sd}^k[m_1] {\boldsymbol C}_k {\boldsymbol h}_{sd,k}[m_1]b_k[i] \\
  \sum_{k=1}^K  a_{{r_1}d}^k[m_2] {\boldsymbol C}_k {\boldsymbol h}_{{r_1}d,k}[m_2]{\tilde b}_k^{{r_1}d}[i] \\
  \vdots \\
  \sum_{k=1}^K  a_{{r_{n_r}}d}^k[m_{n_p}] {\boldsymbol C}_k {\boldsymbol h}_{r_{n_r}d,k}[m_{n_p}]{\tilde b}_k^{{r_{n_r}}d}[i]
\end{array} \hspace{-0.5em} \right] \\ & \quad + {\boldsymbol \eta}[i] + {\boldsymbol n}[i]
\end{split}
\end{equation}}
Rewriting the above signals in a compact form yields
\begin{equation}
\begin{split}
{\boldsymbol r}[i] & = \sum_{k=1}^{K}  \widetilde{\boldsymbol
B}_k[i] \widetilde{\boldsymbol A}_k[i]
\underbrace{\widetilde{\boldsymbol {\mathcal C}}_k {\boldsymbol
h}_k[i]}_{{\boldsymbol p}_k[i]}+ {\boldsymbol \eta}[i] +
{\boldsymbol n}[i]
\\ & = \sum_{k=1}^{K}   \widetilde{\boldsymbol
B}_k[i] \widetilde{\boldsymbol A}_k[i] \widetilde{\boldsymbol
{\mathcal C}}_k {\boldsymbol  h}_k[i]+ {\boldsymbol \eta}[i] +
{\boldsymbol n}[i]
\\ & = \sum_{k=1}^{K}  {\boldsymbol P}_k[i]
{\boldsymbol B}_k[i] {\boldsymbol a}_k[i]+ {\boldsymbol \eta}[i] +
{\boldsymbol n}[i] 
, \label{recdata}
\end{split}
\end{equation}
where the $(n_r+1)M \times (n_r+1)L$ matrix
$\widetilde{\boldsymbol {\mathcal C}}_k = {\rm diag} \{
{\boldsymbol C}_k \ldots {\boldsymbol C}_k \}$ contains copies of
${\boldsymbol C}_k$ shifted down by $M$ positions for each group of $L$ columns and zeros elsewhere. 
The $(n_r+1)L \times 1$ vector ${\boldsymbol h}_k[i]$ contains the
channel gains of the links between the source, the relays and the
destination, and ${\boldsymbol p}_k[i]$ is the effective signature
for user $k$. The $(n_r+1) \times (n_r+1)$ diagonal matrix
${\boldsymbol B}_k[i] = {\rm diag}(b_k[i]~ {\tilde
b}_k^{{r_1}d}[i] \ldots {\tilde b}_k^{{r_n}d}[i]) $ contains the
symbols transmitted from the source to the destination ($b_k[i]$)
and the $n_r$ symbols transmitted from the relays to the
destination (${\tilde b}_k^{{r_1}d}[i] \ldots {\tilde
b}_k^{{r_n}d}[i]$) on the main diagonal, and the $(n_r+1)M \times
(n_r+1)M$ diagonal matrix $\widetilde{\boldsymbol B}_k[i] = {\rm
diag}(b_k[i]\bigotimes {\boldsymbol I}_M~ {\tilde
b}_k^{{r_1}d}[i]\bigotimes {\boldsymbol I}_M \ldots {\tilde
b}_k^{{r_n}d}[i]\bigotimes {\boldsymbol I}_M)$, where $\bigotimes$
denotes the Kronecker product and ${\boldsymbol I}_M$ is an
identity matrix with dimension $M$. The $(n_r+1) \times 1$ power
allocation vector ${\boldsymbol
a}_k[i]=[a_{sd}^k[m_1]~a_{{r_1}d}^k[m_2]\ldots
a_{{r_{n_r}}d}^k[m_{n_p}]]^T$ has the amplitudes of the links, the
$(n_r+1) \times (n_r+1)$ diagonal matrix ${\boldsymbol A}_k[i]$ is
given by ${\boldsymbol A}_k[i] = {\rm diag} \{ {\boldsymbol
a}_k[i] \}$, and the $(n_r+1)M \times (n_r+1)M$ diagonal matrix
$\widetilde{\boldsymbol A}_k[i]= [a_{sd}^k[m_1]\bigotimes
{\boldsymbol I}_M~a_{{r_1}d}^k[m_2]\bigotimes {\boldsymbol I}_M
\ldots a_{{r_{n_r}}d}^k[m_{n_p}]\bigotimes {\boldsymbol I}_M]^T $.
The $(n_r+1)M \times (n_r+1)$ matrix ${\boldsymbol P}_k$ has
copies of the effective signature ${\boldsymbol p}_k[i]$ shifted
down by $M$ positions for each column and zeros elsewhere. The
$(n_r+1)M \times 1$ vector ${\boldsymbol \eta}[i]$ represents the
ISI terms and the $(n_r+1)M \times 1$ vector ${\boldsymbol n}[i]$
has the noise components.

\section{Proposed MMSE Receiver Design, Power Allocation and Channel Estimation}

In this section, a joint receiver design and power allocation
strategy is proposed using constrained linear MMSE estimation and
group-based power constraints along with a linear MMSE channel
estimator. To this end, the $(n_r+1)M \times 1$ received vector in
(\ref{recdata}) can be expressed as
\begin{equation}
{\boldsymbol r}[i] = {\boldsymbol P}_{\mathbf {\mathcal S}}[i]
{\boldsymbol B}_{\mathbf {\mathcal S}}[i] {\boldsymbol
a}_{{\mathbf {\mathcal S}},k}[i] + \sum_{k \neq {\mathbf {\mathcal
S}}} {\boldsymbol P}_k[i] {\boldsymbol B}_k[i] {\boldsymbol
a}_k[i]+ {\boldsymbol \eta}[i] + {\boldsymbol n}[i],
\label{recdatag}
\end{equation}
where ${\mathbf {\mathcal S}} = \{{\mathcal S}_1, {\mathcal S}_2,
\ldots, {\mathcal S}_G \}$ denotes the group of $G$ users to
consider in the design. The $(n_r+1)M \times G(n_r+1)$ matrix
${\boldsymbol P}_{\mathbf {\mathcal S}} = [ {\boldsymbol
P}_{{\mathcal S}_1} ~ {\boldsymbol P}_{{\mathcal S}_2} ~ \ldots ~
{\boldsymbol P}_{{\mathcal S}_G} ]$ contains the $G$ effective
signatures of the group of users. The $G(n_r+1) \times G(n_r+1)$
diagonal matrix ${\boldsymbol B}_{\mathbf {\mathcal S}}[i] = {\rm
diag}(b_{{\mathcal S}_1}[i]~ {\tilde b}_{{\mathcal
S}_1}^{{r_1}d}[i] \ldots {\tilde b}_{{\mathcal S}_1}^{{r_n}d}[i]~
\ldots ~ b_{{\mathcal S}_G}[i]~ {\tilde b}_{{\mathcal
S}_G}^{{r_1}d}[i] \ldots {\tilde b}_{{\mathcal S}_G}^{{r_n}d}[i])$
contains the symbols transmitted from the sources to the
destination and from the relays to the destination of the $G$
users in the group on the main diagonal, the $G(n_r+1) \times 1$
power allocation vector ${\boldsymbol a}_{{\mathbf {\mathcal
S}},k}[i]=[a_{sd}^{{\mathcal S}_1}[i]~a_{{r_1}d}^{{\mathcal
S}_1}[i] \ldots a_{{r_{n_r}}d}^{{\mathcal S}_1}[i],~ \ldots, ~
a_{sd}^{{\mathcal S}_G}[i]~a_{{r_1}d}^{{\mathcal S}_G}[i] \ldots
a_{{r_{n_r}}d}^{{\mathcal S}_G}[i]]^T$ of the amplitudes of the
links used by the $G$ users in the group.

\subsection{Linear MMSE Receiver Design and Power Allocation Scheme with
Group-Based Constraints}

The linear MMSE interference suppression for user $k$ is performed
by the receive filter ${\boldsymbol w}_k[i]=[ {w}_{k,1}[i],~
\ldots, ~ {w}_{k,(n_r+1)M}[i]]$ with $(n_r+1)M$ coefficients on
the received data vector ${\boldsymbol r}[i]$ and yields
\begin{equation}
z_k[i] = {\boldsymbol w}_k^H[i] {\boldsymbol r}[i],
\end{equation}
where $z_k[i]$ is an estimate of the symbols, which are processed
by a slicer $Q(\cdot)$ that performs detection and obtains the
desired symbol as $\hat{b}_k[i] = Q (z_k[i])$.

Let us now detail the linear MMSE-based design of the receivers
for user $k$ represented by ${\boldsymbol w}_k[i]$ and for the
computation of the $G(n_r +1) \times 1$ power allocation vector
${\boldsymbol a}_{{\mathbf {\mathcal S}},k}[i]$. This problem can
be cast as
\begin{equation}
\begin{split}
[ {\boldsymbol w}_{k}^{\rm opt}, ~{\boldsymbol a}_{{{\mathbf
{\mathcal S}},k}}^{\rm opt}  ] & = \arg \min_{{\boldsymbol
w}_k[i], {\boldsymbol a}_{{\mathbf {\mathcal S}},k}[i]} ~
E[ (|{ b}_k[i] - {\boldsymbol w}^H_k[i]{\boldsymbol r}[i] |^2  ] \\
& {\rm subject ~to~}  {\boldsymbol a}_{{\mathbf {\mathcal
S}},k}^H[i] {\boldsymbol a}_{{\mathbf {\mathcal S}},k}[i] = P_{G},
\label{probg}
\end{split}
\end{equation}
The MMSE expressions for the receive filter ${\boldsymbol
w}_{k}[i]$ and the power allocation vector ${\boldsymbol
a}_{{\mathbf {\mathcal S}},k}[i]$ can be obtained by employing
the method of Lagrange multipliers with (\ref{probg}), which leads
to {\small
\begin{equation}
\begin{split}
{\mathcal L}_k & = E\big[ |b_k[i] - {\boldsymbol w}_k^H[i] \big(
{\boldsymbol P}_{\mathbf {\mathcal S}}[i] {\boldsymbol B}_{\mathbf
{\mathcal S}}[i] {\boldsymbol a}_{{\mathbf {\mathcal S}},k}[i] \\
& \quad  + \sum_{k \neq {\mathbf {\mathcal S}}} {\boldsymbol
P}_k[i] {\boldsymbol B}_k[i] {\boldsymbol a}_k[i]   + {\boldsymbol
\eta}[i] + {\boldsymbol n}[i]\big) |^2 \big]   \\
& \quad + \lambda_k ({\boldsymbol a}_{{\mathbf {\mathcal S}},k}[i]
- P_{G}) , \label{lagt}
\end{split}
\end{equation}}
where $\lambda_k$ is a Lagrange multiplier. An expression for
${\boldsymbol a}_{{\mathbf {\mathcal S}},k}[i]$ is obtained by
fixing ${\boldsymbol w}_k[i]$, taking the gradient terms of the
Lagrangian and equating them to zero, which yields
\begin{equation}
{\boldsymbol a}_{{\mathbf {\mathcal S}},k}[i] = ( {\boldsymbol
R}_{{\mathbf {\mathcal S}},k}[i] + \lambda_k {\boldsymbol I})^{-1}
{\boldsymbol p}_{{\mathbf {\mathcal S}},k}[i] \label{avect}
\end{equation}
where the $G(n_r+1) \times G(n_r+1)$ covariance matrix
${\boldsymbol R}_{{\mathbf {\mathcal S}},k}[i] = E[ {\boldsymbol
B}_{\mathbf {\mathcal S}}^H[i]{\boldsymbol P}_{\mathbf {\mathcal
S}}^H[i]  {\boldsymbol w}_k[i] {\boldsymbol w}^H_k[i]{\boldsymbol
P}_{\mathbf {\mathcal S}}[i] {\boldsymbol B}_{\mathbf {\mathcal
S}}[i]]$  and the vector ${\boldsymbol p}_{{\mathbf {\mathcal
S}},k}[i] = E[b_k[i] {\boldsymbol B}_{\mathbf {\mathcal
S}}^H[i]{\boldsymbol P}_{\mathbf {\mathcal S}}^H[i]  {\boldsymbol
w}_k[i]]$ is a $G(n_r+1) \times 1$ cross-correlation vector. The
Lagrange multiplier $\lambda_k$ plays the role of a regularization
term and has to be determined numerically due to the difficulty of
evaluating its expression. Now fixing ${\boldsymbol a}_{{\mathbf
{\mathcal S}},k}[i]$, taking the gradient terms of the Lagrangian
and equating them to zero leads to
\begin{equation}
{\boldsymbol w}_k[i] = {\boldsymbol R}^{-1}[i] {\boldsymbol
p}_k[i], \label{wvect}
\end{equation}
where the covariance matrix of the received vector is given by
${\boldsymbol R}[i] = E[{\boldsymbol r}[i]{\boldsymbol r}^H[i]]$
and ${\boldsymbol p}_k[i] = E[b_k^*[i] {\boldsymbol r}[i]] $ is a
$(n_r+1)M \times 1$ cross-correlation vector. The quantities
${\boldsymbol R}[i]$ and ${\boldsymbol p}_k[i]$ depend on the
power allocation vector ${\boldsymbol a}_{{\mathbf {\mathcal
S}},k}[i]$. The expressions in (\ref{avect}) and (\ref{wvect}) do
not have a closed-form solution as they have a dependence on each
other. Moreover, the expressions also require the estimation of
the channel vector ${\boldsymbol h}_k[i]$. Thus, it is necessary
to iterate (\ref{avect}) and (\ref{wvect}) with initial values to
obtain a solution and to estimate the channel. The network has to
convey the information from the group of users which is necessary to
compute the group-based power allocation including the filter
${\boldsymbol w}_k[i]$. The expressions in (\ref{avect}) and
(\ref{wvect}) require matrix inversions with cubic complexity (
$O(((n_r+1)M)^3)$ and $O((K(n_r+1))^3)$.

\subsection{Cooperative MMSE Channel Estimation}

In order to estimate the channel in the cooperative system
under study, let us consider the transmitted signal for user $k$,
${\boldsymbol x}_{k}[i] = \widetilde{\boldsymbol B}_k[i]
\widetilde{\boldsymbol A}_k[i] \widetilde{\boldsymbol {\mathcal
C}}_k {\boldsymbol h}_k[i]= {\boldsymbol Q}_k[i]{\boldsymbol
h}_k[i]$, and the covariance matrix given by
\begin{equation}
\begin{split}
{\boldsymbol R}& =[{\boldsymbol r}[i] {\boldsymbol r}^H[i]] \\ & =
\sum_{k=1}^{K} {\boldsymbol Q}_k[i] E[ {\boldsymbol h}_k[i]
{\boldsymbol h}_k^H[i]]{\boldsymbol Q}_k^H[i] + E[{\boldsymbol
\eta}[i] {\boldsymbol \eta}^H[i]] + \sigma^2 {\boldsymbol I} \\ &
= \sum_{k=1}^{K} {\boldsymbol Q}_k[i] {\boldsymbol
P}_{{\boldsymbol h}_k} {\boldsymbol Q}_k^H[i] + {\boldsymbol
P}_{\eta} + \sigma^2 {\boldsymbol I}
\end{split}
\end{equation}
A linear estimator of ${\boldsymbol h}_k[i]$ applied to
${\boldsymbol r}[i]$ can be represented as $\hat{\boldsymbol
h}_k[i] = {\boldsymbol T}^H_k{\boldsymbol r}[i]$. The linear MMSE
channel estimation problem for the cooperative system under
consideration is formulated as
\begin{equation}
\begin{split}
{\boldsymbol T}_{k,{\rm opt}} & = \arg \min_{{\boldsymbol T}_k} E
\big[ ||{\boldsymbol h}_k[i] - \hat{\boldsymbol h}_k[i] ||^2 \big]
\\ & = \arg \min_{{\boldsymbol T}_k} E
\big[ ||{\boldsymbol h}_k[i] - {\boldsymbol T}^H_k{\boldsymbol
r}[i] ||^2 \big].
\end{split}
\end{equation}
Computing the gradient terms of the argument and equating them to
zero yields the MMSE solution
\begin{equation}
\begin{split}
{\boldsymbol T}_{k,{\rm opt}} & = {\boldsymbol R}^{-1}
{\boldsymbol P}_{k},
\end{split}
\end{equation}
where ${\boldsymbol P}_{k} = E[{\boldsymbol r}[i]{\boldsymbol
h}_k^H[i]] =  {\boldsymbol Q}_k[i] E[ {\boldsymbol h}_k[i]
{\boldsymbol h}_k^H[i]]=  {\boldsymbol Q}_k[i] {\boldsymbol
P}_{{\boldsymbol h}_k} $. Using the relation $\hat{\boldsymbol
h}_k[i] = {\boldsymbol T}^H_k{\boldsymbol r}[i]$, we obtain
\begin{equation}
\begin{split}
\hat{\boldsymbol h}_k[i] & = {\boldsymbol T}^H_{k,{\rm
opt}}{\boldsymbol r}[i] = {\boldsymbol P}_{k}^H{\boldsymbol
R}^{-1}{\boldsymbol r}[i]\\ & =  {\boldsymbol P}_{{\boldsymbol
h}_k}^H{\boldsymbol Q}_k^H [i]\big(\sum_{k=1}^{K} {\boldsymbol
Q}_k[i] {\boldsymbol P}_{{\boldsymbol h}_k}{\boldsymbol Q}_k^H[i]
+ {\boldsymbol P}_{\eta} + \sigma^2 {\boldsymbol I}
\big)^{-1}{\boldsymbol r}[i],
\end{split}
\label{cest}
\end{equation}
The expressions in (\ref{cest}) require matrix inversions with
cubic complexity ( $O(((n_r+1)M)^3)$), however, this matrix
inversion is common to (\ref{wvect}) and needs to be computed only
once for both expressions. In what follows, computationally
efficient algorithms with quadratic complexity ($O(((n_r+1)M)^2)$)
based on an alternating optimization strategy will be detailed.

\section{Proposed Adaptive Algorithms}

In this section, we develop adaptive RALS algorithms using a
method to build the group of $G$ users based on the power levels,
and then we employ an alternating optimization strategy for
efficiently estimating the parameters of the receive filters, the
power allocation vectors and the channels. Despite the joint
optimization that is associated with a non-convex problem, the
proposed RALS algorithms have been extensively tested and have not
presented problems with local minima.

The first step in the proposed strategy is to build the group of
$G$ users that will be used for the power allocation and receive
filter design. A RAKE receiver is employed to obtain $z_k^{\rm
RAKE}[i] = (\widetilde{\boldsymbol C}_k\hat{\boldsymbol
h}_k[i])^H{\boldsymbol r}[i]=\hat{\boldsymbol
p}_k^H[i]{\boldsymbol r}[i]$ and the group is formed according to
\begin{equation}
{\rm compute} ~~{\rm the}~~G~~{\rm largest}~~|z_k^{\rm
RAKE}[i]|,~~ k=1,2, \ldots, K. \label{group}
\end{equation}
The design of the RAKE and the other tasks require channel
estimation. The power allocation, receive filter design and
channel estimation expressions given in (\ref{avect}),
(\ref{wvect}) and (\ref{cest}), respectively, are solved by
replacing the expected values with time averages, and RLS-type
algorithms with an alternating optimization strategy. In order to
solve (\ref{cest}) efficiently, we develop a variant of the RLS
algorithm that is described by
\begin{equation}
\hat{\boldsymbol h}_k[i] = \hat{\boldsymbol P}_{{\boldsymbol
h}_k}^H[i] {\boldsymbol Q}_k^H[i] \hat{\boldsymbol R}^{-1}[i]
{\boldsymbol r}[i], \label{cestrec}
\end{equation}
where ${\boldsymbol Q}_k[i] = \widetilde{\boldsymbol B}_k[i]
\widetilde{\boldsymbol A}_k[i] \widetilde{\boldsymbol {\mathcal
C}}_k $, the estimate of the inverse of the covariance matrix
$\hat{\boldsymbol R}^{-1}[i]$ is computed with the matrix
inversion lemma \cite{haykin}
\begin{equation}
{\boldsymbol k}[i] = \frac{\alpha^{-1} \hat{\boldsymbol R}[i-1]
{\boldsymbol r}[i]}{1+\alpha^{-1} {\boldsymbol r}^H[i]
\hat{\boldsymbol R}[i-1] {\boldsymbol r}[i]}\label{kgain},
\end{equation}
\begin{equation}
\hat{\boldsymbol R}[i] = \alpha^{-1} \hat{\boldsymbol R}[i-1] -
\alpha^{-1} {\boldsymbol k}[i] {\boldsymbol r}^H[i]
\hat{\boldsymbol R}[i-1], \label{mil2}
\end{equation}
and
\begin{equation}
\hat{\boldsymbol P}_{{\boldsymbol h}_k}[i] = \alpha
\hat{\boldsymbol P}_{{\boldsymbol h}_k}[i-1] + \hat{\boldsymbol
h}_k[i-1] \hat{\boldsymbol h}_k^H[i-1] \label{Prec},
\end{equation}
where $\alpha$ is a forgetting factor that should be close to but
less than $1$. The approach for allocating the power within a
group is to drop the constraint, estimate the quantities of
interest and then impose the constraint via a subsequent
normalization. The group-based power allocation algorithm
is computed by
\begin{equation}
\begin{split}
\hat{\boldsymbol a}_{{\mathbf{\mathcal S}},k}[i] & =
\hat{\boldsymbol R}_{{\mathbf{\mathcal S}},k}[i] \hat{\boldsymbol
p}_{{\mathbf{\mathcal S}},k}[i]\\ & = \hat{\boldsymbol
R}_{{\mathbf{\mathcal S}},k}[i] (\alpha\hat{\boldsymbol
p}_{{\mathbf{\mathcal S}},k}[i-1] + b_k[i] {\boldsymbol
v}_k[i])\\& =  \hat{\boldsymbol a}_{{\mathbf{\mathcal S}},k}[i-1]
+ \xi_{a}[i] {\boldsymbol k}_{{\mathbf{\mathcal S}},k}[i],
\label{arls}
\end{split}
\end{equation}
where $\xi_a[i] = b_k[i] - \hat{\boldsymbol a}_{{\mathbf{\mathcal
S}},k}^H[i-1] {\boldsymbol v}_k[i]$ is the a priori error,
${\boldsymbol v}_k[i] = {\boldsymbol B}_{\mathbf {\mathcal
S}}^H[i]{\boldsymbol P}_{\mathbf {\mathcal S}}^H[i] {\boldsymbol
w}_k[i]$ is the input signal to the recursion
\begin{equation}
{\boldsymbol k}_{{\mathbf{\mathcal S}},k}[i] = \frac{\alpha^{-1}
\hat{\boldsymbol R}_{{\mathbf{\mathcal S}},k}[i-1] {\boldsymbol
v}_k[i]}{1+\alpha^{-1} {\boldsymbol v}_k^H[i] \hat{\boldsymbol
R}_{{\mathbf{\mathcal S}},k}[i-1] {\boldsymbol v}_k[i] },
\end{equation}
\begin{equation}
\hat{\boldsymbol R}_{{\mathbf{\mathcal S}},k}[i] = \alpha^{-1}
\hat{\boldsymbol R}_{{\mathbf{\mathcal S}},k}[i-1] - \alpha^{-1}
 {\boldsymbol k}_{{\mathbf{\mathcal S}},k}[i] {\boldsymbol
v}_k^H[i] \hat{\boldsymbol R}_{{\mathbf{\mathcal S}},k}[i-1].
\end{equation}
The normalization $\hat{\boldsymbol a}_{{\mathbf{\mathcal
S}},k}[i] \leftarrow P_G  ~\hat{\boldsymbol a}_{{\mathbf{\mathcal
S}},k}[i]/||\hat{\boldsymbol a}_{{\mathbf{\mathcal S}},k}[i]|| $
is then performed to ensure the power constraint. The receive
filter is computed by
\begin{equation}
\hat{\boldsymbol w}_k[i] = \hat{\boldsymbol w}_k[i-1] +
{\boldsymbol k}[i] \xi^*[i], \label{wrls}
\end{equation}
where the a priori error is given by $\xi[i] = b_k[i] -
\hat{\boldsymbol w}_k^H[i-1] {\boldsymbol r}[i]$ and ${\boldsymbol
k}[i]$ is given by (\ref{kgain}). The proposed scheme employs the
algorithm in (\ref{group}) to allocate the users in the group and
the channel estimation approach of (\ref{cestrec})-(\ref{Prec}).
The alternating optimization strategy uses the recursions
(\ref{arls}) and (\ref{wrls}) with $1~{\rm or}~2$ iterations per
symbol $i$.

\section{Simulations}

The bit error ratio (BER) performance of the proposed joint power
allocation and interference suppression (JPAIS) scheme and RALS
algorithms with group-based power constraints (GBC) is assessed. The
JPAIS scheme and algorithms are compared with schemes without
cooperation (NCIS) and with cooperation (CIS) \cite{venturino}
using an equal power allocation across the relays (the power
allocation in the JPAIS scheme is disabled). A DS-CDMA network
with randomly generated spreading codes and a processing gain
$N=16$ is considered. The fading channels are generated
considering a random power delay profile with gains taken from a
complex Gaussian variable with unit variance and mean zero, $L=5$
paths spaced by one chip, and are normalized for unit power. The
power constraint parameter $P_{A,k}$ is set for each user so that
the designer can control the SNR (${\rm SNR} = P_{A,k}/\sigma^2$)
and $P_T= P_G + (K-G) P_{A,k}$, whereas it follows a log-normal
distribution for the users with associated standard deviation
equal to $3$ dB. The DF cooperative protocol is adopted and all
the relays and the destination terminal use either linear MMSE,
which have full channel and noise variance knowledge, or adaptive
receivers. The receivers are adjusted with the proposed RALS with
$2$ iterations for the JPAIS scheme, and with RLS algorithms for
the NCIS and CIS schemes. We employ packets with $1500$ QPSK
symbols and average the curves over $1000$ runs. For the adaptive
receivers, we provide training sequences with $N_{\rm tr}=200$
symbols placed at the preamble of the packets. After the training
sequence, the adaptive receivers are switched to decision-directed
mode.

\begin{figure}[!htb]
\begin{center}
\def\epsfsize#1#2{1\columnwidth}
\epsfbox{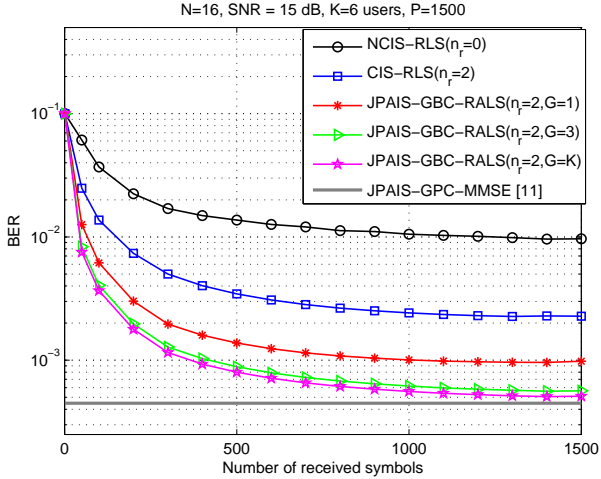} \vspace{-1.5em} \caption{\footnotesize BER
performance versus number of symbols. Parameters:
$\lambda_T=\lambda_k =0.025$ (for MMSE schemes), $\alpha=0.998$,
$\hat{\boldsymbol R}_{{\mathbf{\mathcal S}},k}^{-1}[i]=0.01
{\boldsymbol I}$ and $\hat{\boldsymbol R}^{-1}[i]=0.01 {\boldsymbol
I}$.}  \label{fig1}
\end{center}
\end{figure}

The first experiment depicted in Fig. \ref{fig1} shows the BER
performance of the proposed JPAIS scheme and algorithms against
the NCIS and CIS schemes with $n_r=2$ relays. The JPAIS scheme is
considered with the group-based power constraints (JPAIS-GBC). All
techniques employ MMSE or RLS-type algorithms for estimation of
the channels, the receive filters and the power allocation for
each user. The results show that as the group size $G$ is
increased the proposed JPAIS scheme and algorithms converge to
approximately the same level of the cooperative JPAIS-MMSE scheme
reported in \cite{delamare_jpais}, which employs $G=K$ for power
allocation, and has full knowledge of the channel and the noise
variance.

\begin{figure}[!htb]
\begin{center}
\def\epsfsize#1#2{1\columnwidth}
\epsfbox{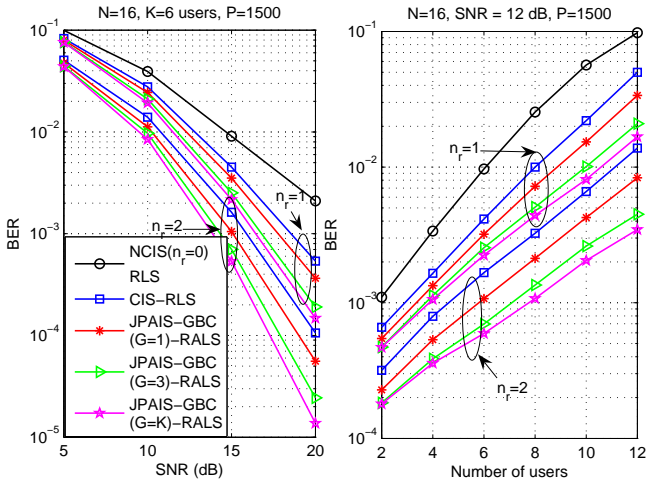} \vspace{-1.5em}\caption{\footnotesize BER
performance versus SNR and number of users for the optimal linear
MMSE detectors. Parameters: $\alpha =0.998$, $\hat{\boldsymbol
R}_{{\mathbf{\mathcal S}},k}^{-1}[i]=0.01 {\boldsymbol I}$ and
$\hat{\boldsymbol R}^{-1}[i]=0.01 {\boldsymbol I}$.}  \label{fig2}
\end{center}
\end{figure}

The proposed JPAIS-GBC scheme is then compared with a
non-cooperative approach (NCIS) and a cooperative scheme with
equal power allocation (CIS) across the relays for $n_r=1,2$
relays. The results shown in Fig. \ref{fig2} illustrate the
performance improvement achieved by the JPAIS scheme and
algorithms, which significantly outperform the CIS and the NCIS
techniques. As the number of relays is increased so is the
performance, reflecting the exploitation of the spatial diversity.
In the scenario studied, the proposed JPAIS-GBC with $G=3$ can
accommodate up to $3$ more users as compared to the CIS scheme and
double the capacity as compared with the NCIS for the same BER
performance. The curves indicate that the GBC for power allocation
with only a few users is able to attain a performance close to the
JPAIS-GBC with $G=K$ users, while requiring a lower complexity and
less network signalling. A comprehensive study of the signalling
requirements will be considered in a future work.

\section{Concluding Remarks}

This work has proposed the JPAIS scheme with group-based
constraints (GBC) for cooperative DS-CDMA networks with multiple
hops and the DF protocol. A constrained MMSE design for the
receive filters and the power allocation with GBC
has been devised along with an MMSE channel estimator. We have
proposed RALS algorithms for estimating the parameters of the
channels, the receive filter and the power allocation. The results
have shown that the JPAIS scheme with GBC and the RALS
algorithms achieve significant gains in performance and capacity
over existing schemes. 

\end{document}